\documentclass[conference]{IEEEtran} 
\usepackage{amsmath,amssymb,amsfonts}
\usepackage{cite}
\usepackage{bm}
\usepackage{mathtools}
\usepackage[binary-units=true]{siunitx}
\DeclareSIUnit{\belmilliwatt}{Bm}
\usepackage{graphicx}
\usepackage{dsfont}
\usepackage{subcaption}
\usepackage{comment}
\usepackage{algorithm}
\usepackage{algpseudocode}
\usepackage{mathtools}

\usepackage{textcomp}
\usepackage[export]{adjustbox}
\usepackage{float}
\usepackage{booktabs}
\usepackage{multicol}
\usepackage{xparse}
\usepackage{color,soul}
\usepackage[bottom]{footmisc}
\usepackage[binary-units=true]{siunitx}
 \usepackage{bm}           
\usepackage{tikz}         
\usetikzlibrary{arrows.meta, positioning, calc}
\DeclareSIUnit{\belmilliwatt}{Bm}
\DeclareSIUnit{\dBm}{\deci\belmilliwatt}
\DeclareSIUnit[per-mode=symbol,per-symbol=p]{\Bps}{\byte\per\second}

\sethlcolor{yellow}
\usepackage{algpseudocode}

\usepackage{hyperref}

\makeatletter
\def\BState{\State\hskip-\ALG@thistlm}
\makeatother

\IEEEoverridecommandlockouts
\begin{document}

    \title{Distributed Online Learning  for Time-Critical Communication in 6G Industrial Subnetworks}

\author{
\IEEEauthorblockN{Samira Abdelrahman\IEEEauthorrefmark{1}, Hossam Farag\IEEEauthorrefmark{2}\IEEEauthorrefmark{1}, and  Gilberto Berardinelli\IEEEauthorrefmark{2}}
\IEEEauthorblockA{
\IEEEauthorrefmark{1} Department of Electrical Engineering, Aswan University, Egypt\\
\IEEEauthorrefmark{2}Department of Electronic Systems, Aalborg University, Denmark \\
Email: sma@asw.edu.eg,  \{hmf, gb\}@es.aau.dk}
}

	\maketitle
	
	\begin{abstract}
Reliable and timely delivery of emergency alarms is critical for 6G industrial in-X subnetworks, particularly in dense and dynamic environments where multiple devices may simultaneously detect the same event and contend for limited radio resources. This scenario, known as medium access with a shared message, poses significant challenges for conventional random access protocols due to collisions and stringent transmission deadlines. To address this challenge, this paper proposes a distributed deep contextual-bandit-based random access protocol that minimizes the probability of alarm delivery deadline violations. The proposed protocol enables each local access point (LAP) to autonomously learn its transmission strategy through a lightweight deep neural network (DNN), without requiring knowledge of the number of active LAPs or explicit inter-LAP coordination. A contention-signature acquisition mechanism is introduced to provide contextual information about network contention, enabling distributed and adaptive channel-access decisions. Moreover, an online learning mechanism allows LAPs to continuously refine their transmission strategies based on feedback from previous alarm transmissions. Simulation results demonstrate that the proposed protocol outperforms benchmark random access schemes, particularly under high-contention conditions, improving the probability of timely alarm delivery by at least $8\%$ with $30$ subnetworks and up to $21\%$ with 60 subnetworks.
	\end{abstract}
\section{Introduction}\label{sec:intro}
The advent of sixth-generation (6G) wireless systems is expected to enable a new era of hyper-reliable, low-latency, and intelligent connectivity via the vision of ``network of networks''~\cite{6GG}. At the edge of such a ``network of networks'' architecture, 6G in-X subnetworks are located to provide localized, short-range radio communication within entities such as robots, vehicles and humans~\cite{IN-X1, resources}. In the context of future industrial systems, in-X subnetworks can be installed inside robots and/or production modules, enabling localized wireless control, e.g., control of moving parts in mobile robots~\cite{6Gshine}. 


Mobile robots equipped with in-X subnetworks (short-range, low-power cells) provide a scalable, resilient, and high-performance solution for condition monitoring in critical, large-scale industrial environments. These environments are characterized by high spatial complexity, harsh operating conditions, and safety-critical processes, where failures can rapidly propagate and lead to severe consequences. Deploying dense fixed-sensor infrastructure is often costly, difficult to scale, and prone to coverage gaps in large, hazardous, or dynamically changing environments~\cite{fixed1, fixed2}. In contrast, a fleet of autonomous mobile robots, each equipped with an in-X subnetwork, enables flexible and adaptive monitoring, as sensing, processing, and control are co-located within each entity. Within each subnetwork, sensors and actuators interact through local control loops to perform navigation and autonomous inspection tasks, including close-range measurements enabled by robotic manipulation. During inspection, robots may detect abnormal conditions such as gas leaks, structural defects, or overheating components, triggering the transmission of emergency alarms to a central controller. The response to such emergency events may require the robots (subnetworks) to perform distinct actions synergistically. For example, in the event of a detected gas leak, one robot equipped with a valve actuator may be instructed to shut down a pipeline segment, while another robot carrying a sealing mechanism may physically isolate the leakage point. We focus on a time-critical communication scenario where multiple robots may simultaneously detect the same abnormal event and attempt to report a shared alarm message to a central controller. This scenario is referred to as medium access with a shared message~\cite{NP}. For example when a gas leak spreads across several monitored zones or when a machine failure affects nearby equipment~\cite{6Gshine}. The resulting contention for limited radio resources can lead to collisions and deadline violations, particularly in dense deployments. Addressing this challenge requires a distributed medium access mechanism capable of adapting to varying contention conditions without relying on centralized scheduling or explicit coordination among the transmitting devices.

In this work, we propose a distributed deep contextual-bandit-based random access (RA) protocol that enables local access points (LAPs) to autonomously learn their channel-access strategies from contextual observations and transmission feedback. Unlike conventional RA approaches, the proposed protocol exploits a contention-signature (CS) signal to facilitate implicit coordination among active LAPs. The learning process is performed locally using a lightweight deep neural network (DNN), allowing transmission decisions to adapt to changing contention conditions without requiring prior knowledge of the number or identities of active LAPs. Furthermore, the protocol explicitly accounts for alarm retransmissions within a strict delivery deadline. The main contributions of this work are summarized as follows:
\begin{itemize}
    \item  We develop a deep contextual-bandit-based RA protocol for the shared-message alarm scenario in 6G industrial in-X subnetworks. Each LAP independently learns a transmission policy using a compact DNN to estimate the expected reward associated with different channel-access patterns. The proposed approach supports a varying number of simultaneously active LAPs without requiring explicit inter-LAP coordination or knowledge of the active device population.
    \item We introduce a CS acquisition mechanism in which active LAPs transmit pilot signals that are aggregated at a central access point (CAP) and subsequently broadcast to the LAPs. The resulting CS provides contextual information correlated with the current contention conditions, enabling distributed and informed channel-access decisions. Furthermore, the protocol allows LAPs to independently reselect their transmission patterns following unsuccessful attempts without reacquiring the CS, thereby supporting retransmissions with limited signaling overhead.
\item We develop an online training procedure in which each LAP updates its DNN based on transmission feedback. In addition, we formulate the deadline-constrained alarm delivery problem and employ a discrete-time Markov chain (DTMC) to characterize the relationship between the transmission success probability and the deadline-violation probability.
\item We evaluate the proposed protocol through MATLAB simulations considering different network densities, channel availability, and alarm activation probabilities. Comparisons with benchmark RA schemes demonstrate improved timely alarm delivery, particularly under high-contention conditions. We also investigate the impact of DNN architecture and learning behavior on the performance of the proposed approach.
\end{itemize}
The rest of the paper is organized as follows. Section~\ref{related} introduces the background and related work. 
Section~\ref{system-model} describes the system model and
Section~\ref{proposed} presents the proposed deep contextual-bandit-based RA method.
Section~\ref{results} presents the performance evaluation, and finally
Section~\ref{sec:conclusions} concludes the paper.
\section{Background and Related Work}
\label{related}
In the context of industrial wireless networks, industrial protocols such as WirelessHART, ISA~100.11a, and WIA-PA primarily rely on Time-Division Multiple Access (TDMA) for channel access management~\cite{deadline}. TDMA is effective in providing deterministic latency and eliminating collisions for periodic transmissions. However, it is ill-suited for handling sporadic, event-driven traffic such as emergency alarms, where packets may experience substantial delays waiting for their assigned transmission slots, potentially violating critical latency requirements. Conversely, Random Access (RA) schemes like ALOHA permit unscheduled transmissions, enabling devices to immediately react to sporadic events with low signaling requirements. However, the lack of coordination inherent in RA significantly increases the risk of collisions, especially in scenarios with high device density and synchronous traffic bursts triggered by a common event (e.g., violation of control stability). Importantly, both TDMA and RA were not originally developed considering the demands and  peculiar characteristics of in-X subnetworks, where challenges such as mobility, high density, and stringent real-time constraints are prevalent. Therefore, the envisioned extreme density and large-scale deployment of subnetworks in a factory scenario requires novel scheduling and RA protocols capable of coordinating channel access and mitigating interference in densely deployed industrial networks. In this regard, intelligent RA solutions based on Machine Learning (ML) have emerged as promising alternatives~\cite{RA-ML1, RA-ML2, RA-ML3}. By learning from  past interactions and contextual observations~\cite{sub5}, subnetworks can autonomously and in a distributed manner adapt their transmission behavior to optimize access efficiency under dynamic industrial conditions while reducing the need for explicit signaling.

Different works~\cite{RA2, RA3, NP}  have proposed improved communication protocols to enhance delay and reliability in industrial networks. In~\cite{RA2}, the authors proposed  a multi-agent deep reinforcement learning (DRL) framework for RA in massive IoT environments, adopting a centralized training and decentralized execution paradigm. However, the proposed random access policy fails to converge in scenarios where device identifiers are unavailable, limiting its applicability in decentralized and anonymous industrial deployments. A multi-armed bandit (MAB)-based method is introduced in~\cite{NP} that enables indirect coordination among devices during transmission of time-critical updates. However, the proposed method assumes that no more than three devices are active at the same time, which is impractical in realistic industrial settings where an alarm event could activate many devices simultaneously. Moreover, these works are not primarily designed considering the peculiar nature of short-range low-power in-X subnetworks. In recent years, several works have been introduced targeting interference management and radio resource allocation for in-X subnetworks~\cite{sub3, sub4, sub5, sub6, sub7}. For instance, the authors in~\cite{sub3, sub4} use graph neural networks to model the spatial dependencies in wireless subnetworks and optimize power control policies. Other recent studies~\cite{sub5, sub6, sub7} have employed multi-agent reinforcement learning  to address dynamic spectrum access and resource allocation in mobile and dense in-X subnetworks. However, the resulting learning policies may improve overall network efficiency without necessarily prioritizing urgent, time-critical transmissions. Moreover, these works  primarily target average performance metrics such as throughput, outage probability and energy efficiency, and are not inherently designed to handle deadline-sensitive or alarm-driven traffic, which requires real-time responsiveness. In addition, while these studies focus on improving local communication performance (i.e., sensor-actuator communication), none of them addresses the problem of contention-based reporting of a common time-critical alarm. This introduces a distinct medium access problem in which transmission decisions across subnetworks must be implicitly coordinated to mitigate collisions and maximize the probability of timely alarm delivery.
\section{System Model and Problem Formulation}
\label{system-model}
\subsection{Description of the Use-Case and Architecture of the Subnetworks}
    \begin{figure}[t] 
		\centering
		\includegraphics[width= 1\linewidth]{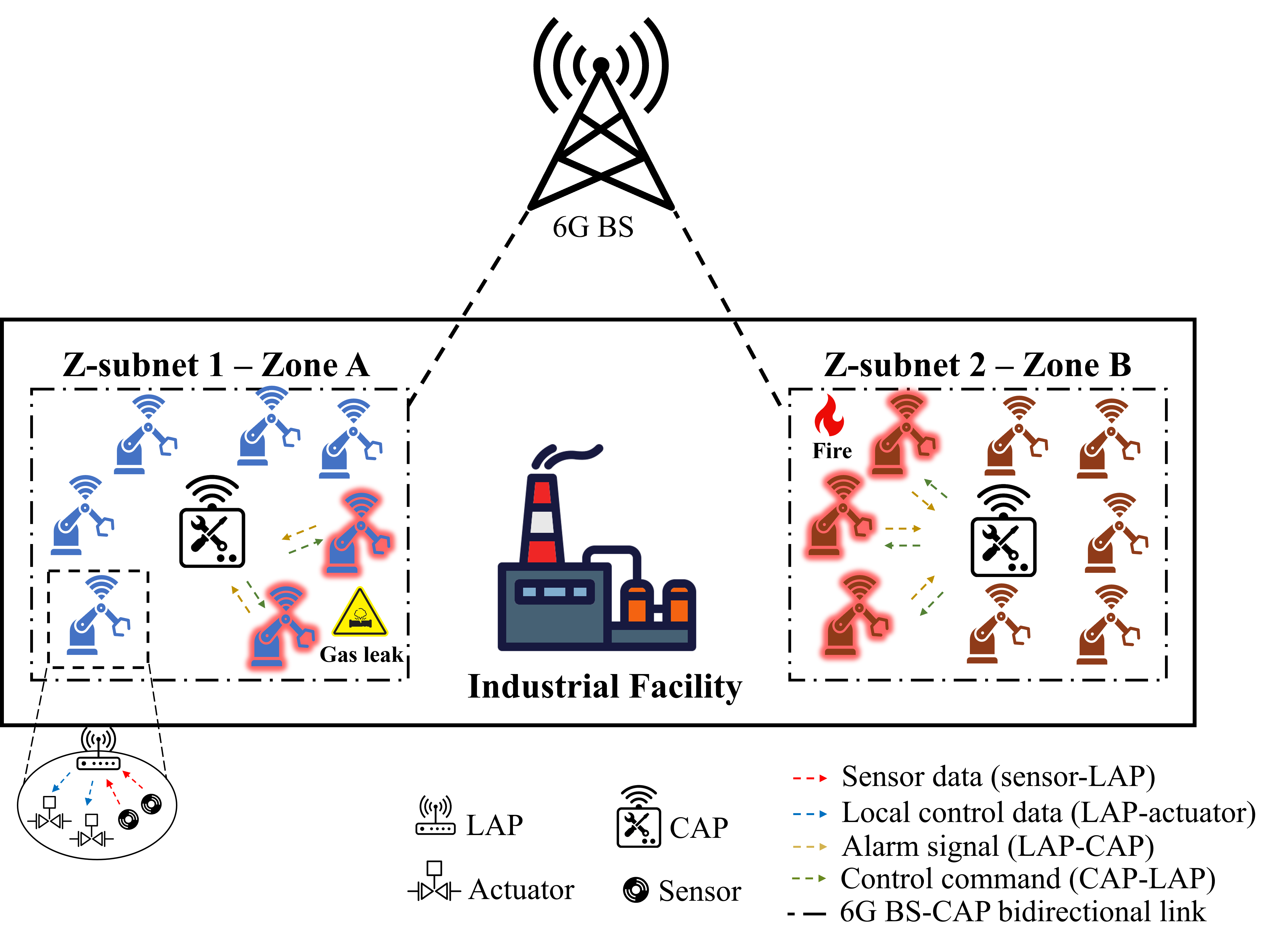}
		\caption{Network model of an industrial scenario comprising a set of Z-subnets connected to a parent 6G network.  \label{system}}
	\end{figure}

 As illustrated in Fig.~\ref{system}, we consider the scenario of subnetwork swarms~\cite{6Gshine} where the industrial facility is partitioned into multiple localized operational zones, each managed by a set of heterogeneous robots engineered to perform distinct sensing and actuation functions. The devices within each zone exchange critical traffic, such as sensor and control information, forming a localized, short-range subnetwork (Z-subnet). Within each Z-subnet, a designated device, equipped with edge-computing and programmable logic controllers (PLCs) capabilities, acts as a CAP. All the Z-subnets are connected to a 6G base station (BS) which functions as the central entity for coordination and radio resource management between the different Z-subnets.

Robots within each Z-subnet constantly collect information about the industrial assets within their confined zone via various embedded sensors (e.g., temperature, vibration, camera data) and communicate with the CAP only when an abnormal event is detected (e.g., gas leakage, abnormal vibration, overheating). Each robot is equipped with a LAP which provides radio as well as edge-computing and local control-logic functionalities (e.g., threshold and rule-based safety logic) to detect abnormal events (e.g., fire or motion detection). The considered architecture represents a form of nested  subnetworks, where each group of devices within a zone is represented by a subnetwork (Z-subnet) on a higher hierarchical level and where each of the robots locally implements another subnetwork (L-subnet) on a lower hierarchical level. Since each L-subnet includes a single LAP, the two terms are used interchangeably throughout the paper. Within a Z-subnet, active robots (those that detect the abnormal event) attempt to transmit an alarm signal to the CAP which processes it internally to determine the required control functions. This may involve sophisticated algorithms that consider different robot functions and their states, environmental parameters, etc. The CAP then sends back control instructions to the relevant robots within the subnetwork, thereby closing the control loop. Upon receiving the control commands, each robot executes specific actuation tasks according to its functional role. 

In this work, we focus on the timely and reliable transmission of alarm signals within each Z-subnet. Particularly, we focus on developing a low-overhead RA protocol that maximizes the probability that an alarm message is successfully delivered to the CAP within its deadline bound from at least one active LAP.

\subsection{Network Model}
We assume a homogeneous set of spatially distributed Z-subnets within the industrial facility. For the rest of the paper, our analysis focuses on an arbitrary Z-subnet comprising a set $\mathcal{N}$ ($|\mathcal{N}|=N$) of L-subnets. Each L-subnet represents a robotic unit that comprises a set of sensors and actuators connected to one LAP which provides radio, as well as simple processing capabilities within the unit. Sensors and actuators within each L-subnet are scheduled via TDMA, avoiding intra-subnetwork interference. For the defined nested structure in the previous section, the 6G parent network (e.g., enterprise base station or gNodeB) is responsible for radio resource management among the different Z-subnets. Particularly, we consider that the Z-subnets are scheduled over orthogonal frequency bands to avoid inter-Z-subnet interference. Therefore, an active LAP $n$ interferes only with active LAPs ($\mathcal{\grave{N}}$) within the same Z-subnet.

An abnormal event triggers a random subset $\mathcal{\grave{N}}\subseteq \mathcal{N}$ of LAPs to transmit the alarm information to the CAP. Let $p_{d_n}$ denote the activation probability that quantifies the influence  of the $i$-th event, originating at epicenter $(x_i, y_i)$, on the $n$-th LAP in the two-dimensional Euclidean plane $\mathbb{R}^2$. The variable $d_{i,n}$ represents the Euclidean distance between the event epicenter and the $n$-th LAP. In this context, $p_{d_n}$ represents the likelihood that an LAP successfully detects an event occurring at distance $d_{i,n}$. In practice, $p_{d_n}$ is influenced by the normalized measurement output of the sensors within the L-subnet in response to the event. For instance, the LAP calculates the normalized (against a known maximum expected value) temperature, vibration amplitude, or gas concentration values collected by the sensors within the L-subnet and compares them with a specific threshold. The function $p_{d_n} \in [0,1]$ is assumed to be decreasing with respect to distance, thereby modeling the natural attenuation of event impact as $d_{i,n}$ increases. Several functional forms may be adopted for $p_{d_n}$~\cite{linear, exp, piece-wise}. For analytical tractability, we approximate the distance-dependent detection probability using an exponential decay function~\cite{exp}, which captures the decreasing influence of an event with increasing distance from its epicenter~\cite{spatial}. Therefore, $p_{d_n}$ can be expressed as
\begin{equation}
p_{d_n} = e^{-\eta d_{i,n}},
\end{equation}
where the parameter $\eta > 0$ controls the spatial attenuation rate and therefore determines the average sensing sensitivity as a function of distance.  The value of $\eta$ can be derived from domain-specific constants. 
\subsection{Problem Formulation}
A timely response to an emergency event is crucial and mainly depends on the CAP receiving the alarm information within a predefined deadline $T_D$. Throughout the paper, we use the normalized deadline $D=\frac{T_D}{T_C}$, where $T_C$ is the time slot duration. The LAPs are time-synchronized and share a bandwidth of $M$ orthogonal channels, with $M\ll N$, for alarm transmission to the CAP. Within each L-subnet, local (sensor-actuator) communication is managed independently using TDMA, whereas the proposed RA protocol governs contention among active LAPs sharing the alarm-reporting channels. Consequently, the two communication types (intra and inter-L-subnet) coexist without mutual interference. The goal is to ensure that the CAP successfully receives the alarm within $D$ time slots on at least one channel. Exceeding the deadline means that the CAP will fail to react in a timely manner to the generated alarm.

To transmit the alarm, each activated LAP $n\in \mathcal{\grave{N}}$  selects a transmission pattern $\boldsymbol{b}_n=[b_{n,1}, b_{n,2}, ..., b_{n,M}]^T$ at each transmission attempt, where $b_{n,m}\in \{0,1\}$ is the channel selection indicator
\begin{equation}
    b_{n, m} =
\begin{cases}
1, & \text{if LAP } n \text{ decides to transmit over channel } m, \\
0, & \text{otherwise}.
\end{cases}
\end{equation}
Let the matrix $\boldsymbol{B}\in\{0,1\}^{M\times |\mathcal{\grave{N}}|}$ represent the access configurations selected by the active LAPs. Each column in $\boldsymbol{B}$ matches with one vector in the set $\{\boldsymbol{b}_n|n\in \mathcal{\grave{N}}\}$. A successful reception of the  alarm by the CAP is denoted by the indicator $\delta(\mathcal{\grave{N}}, \boldsymbol{B})$ which is given as
\begin{equation}\label{sucess-ind}
\delta(\mathcal{\grave{N}}, \boldsymbol{B})= I\left(\exists m \in \{1, ..., M\}: \sum_{n \in \mathcal{\grave{N}}} b_{n,m} = 1\right),
\end{equation}
where $I(.)$ is the indicator function, that is equal to 1 when the condition is satisfied. Note that for $\delta(\mathcal{\grave{N}}, \boldsymbol{B})$ in \eqref{sucess-ind}, we consider a simple collision channel, i.e., a transmission fails only due to collision, which is a commonly adopted assumption~\cite{NP}. This assumption isolates the RA problem, which is the focus of this work. Incorporating channel impairments (path loss, shadowing, fading and noise) in the analysis is left for future work. With a total bandwidth of $M$ channels, each LAP $n$ can select from $2^M$ access configurations. All possible access configurations can be defined by the  matrix $\grave{\boldsymbol{B}}=[\grave{\boldsymbol{b_1}}, \grave{\boldsymbol{b_2}}, ..., \grave{\boldsymbol{b_{2^M}}}]\in \{0,1\}^{M\times 2^M}$, where each column represents a unique access configuration, i.e.,  $\grave{\boldsymbol{b_i}}\neq \grave{\boldsymbol{b_j}} \forall i\neq j$. To clarify, assume that we have $|\mathcal{\grave{N}}|=5$ and $M=2$. LAPs 1, 2, and 4 choose to transmit on channel 2,  LAP 5 chooses to transmit on both channels and LAP 3 remains silent. Then, we have
\begin {equation*}
\grave{\boldsymbol{B}} =
\begin{bmatrix}
0 & 0 & 1 & 1 \\
0 & 1 & 0 & 1
\end{bmatrix}
\mathrm{and\,\,}
\boldsymbol{B} =
\begin{bmatrix}
0 & 0 & 0 & 0 & 1 \\
1 & 1 & 0 & 1 & 1
\end{bmatrix}
.
\end{equation*}
We denote the probability that an active LAP $n$ chooses an access configuration $\grave{\boldsymbol{b_i}}$ as  $\psi_{n}(\grave{\boldsymbol{b_i}})$ where we have $\sum_{i=1}^{2^M} \psi_{n}(\grave{\boldsymbol{b_i}}) = 1, \forall n \in \mathcal{\grave{N}}$. In addition, we define the matrix $\boldsymbol{\Psi}$, of size $|\mathcal{N}|\times 2^M$, whose $n^{\textrm{th}}$ row represents the elements of the set $\{\psi_{n}(\grave{\boldsymbol{b_i}})|\forall i=1, ..., 2^M\}$. For all possible $\mathcal{\grave{N}}$, the successful transmission probability is given by 
\begin{align}\label{success_prob}
P_s^{\mathrm{pkt}}(\Psi) =&\sum_{\acute{\mathcal N}\in\mathcal P(\mathcal N)}\Bigg[\prod_{n\in\acute{\mathcal N}}p_{d_n}\prod_{n\in\mathcal N\setminus\acute{\mathcal N}}(1-p_{d_n})\Bigg]\nonumber\\
&\times\sum_{\boldsymbol B\in\{0,1\}^{M\times|\acute{\mathcal N}|}}
\Bigg(\delta(\acute{\mathcal N},\boldsymbol B)\prod_{n\in\acute{\mathcal N}}\psi_n(\boldsymbol b_n)\Bigg),
\end{align}
    \begin{figure}[t] 
		\centering
		\includegraphics[width=1\linewidth]{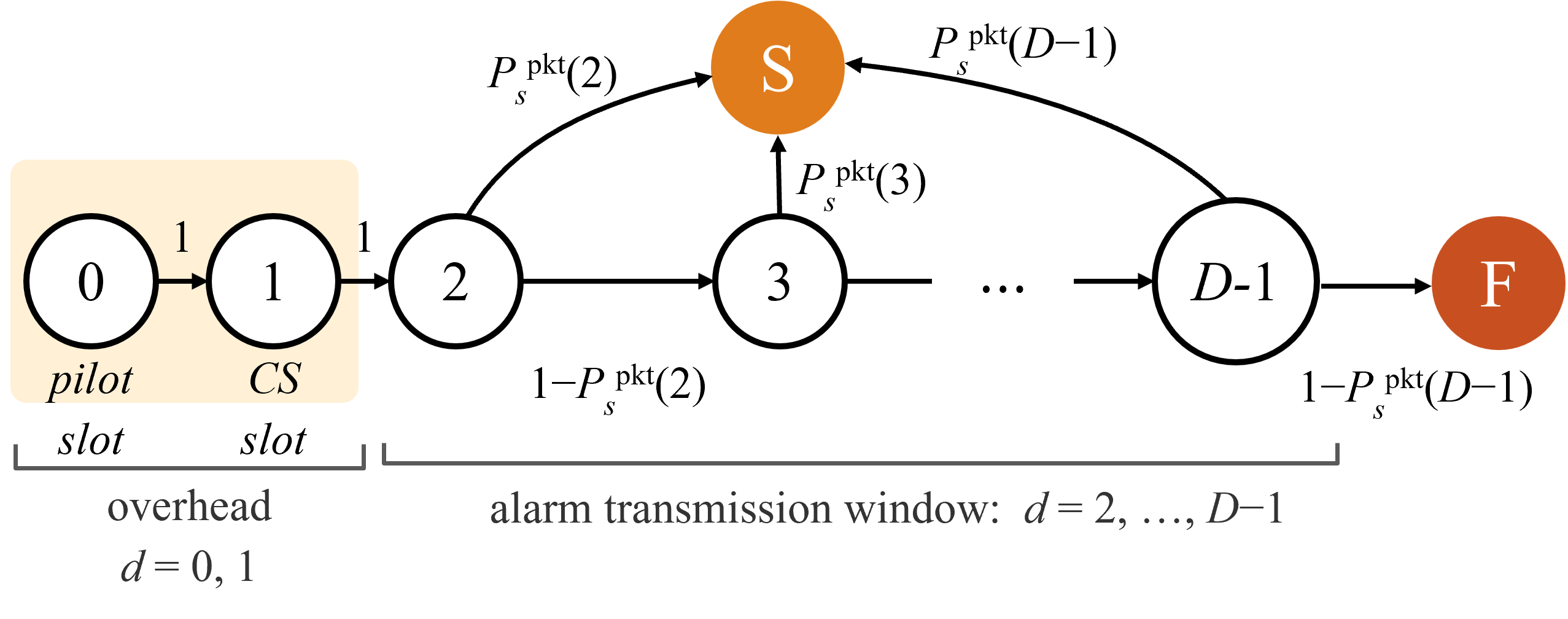}
		\caption{DTMC model for a tagged alarm packet with two overhead slots.} 
        \label{DTMC-deadline}
	\end{figure}

Next, we calculate the probability that the alarm packet is delivered within its deadline of $D$ slots. Define the packet age state $d\in\{0,1,\dots,D-1\}$ as the number of elapsed slots since alarm generation. Each alarm event follows a fixed slot structure. In the first slot after the
event (slot~0), every active LAP transmits its pilot sequence to the CAP. In the
second slot (slot~1), the CAP broadcasts the aggregated signal, i.e., the CS
signal. These two slots are protocol overhead: no alarm transmission can take
place in them, and they are not available for retransmissions. Alarm
transmissions can therefore start only in slot~2. Since the deadline $D$ is
measured from the instant the alarm is generated, only $D-2$ slots
(slots $2,\dots,D-1$) remain for delivering the alarm.
We construct a discrete-time Markov chain (DTMC) (see Fig.~\ref{DTMC-deadline}) with transient states $0,1,\dots,D-1$ and two absorbing states: S (success: packet received) and F (failure: deadline missed). Let $X(t)$ denote the chain state at slot boundary $t$ with $X(0)=0$. A successful transmission is recognized by receiving a shared acknowledgment (ACK) message from the CAP over a dedicated error-free channel. For $d=2, \cdots, D-1$, we have
\[
P_s^{\mathrm{pkt}}(d)\triangleq \Pr\big(\text{packet delivered} \mid \text{current age}=d\big),
\]
which, under stationary contention process during the lifetime of the packet, reduces $P_s^{\mathrm{pkt}}(\Psi)$ defined in~\eqref{success_prob}.

Let $p_{d,j}\triangleq\Pr\{X(t{+}1)=j\mid X(t)=d\}$ denote the one-step
transition probability from age state $d$ to state $j$. The pilot and CS
slots ($d\in\{0,1\}$) are traversed deterministically, i.e., $p_{d,d+1}=1$. For $d=2,\dots,D-2$, we have $p_{d,\mathsf{S}} = P_s^{\mathrm{pkt}}(d)$ and $p_{d,d+1}= 1-P_s^{\mathrm{pkt}}(d).$
Finally, for the last transient state $d=D-1$, an unsuccessful attempt means that
the deadline has expired, i.e., $p_{D-1,\mathsf{S}}= P_s^{\mathrm{pkt}}(D-1)$ and $  p_{D-1,\mathsf{F}}= 1-P_s^{\mathrm{pkt}}(D-1).$

Let $P_{>D}$ denote the deadline violation probability  conditioned on the initial age $X(0)=0$. Since the chain traverses
the two overhead states with probability one, the transitions above give
\begin{equation}
P_{>D}=\prod_{i=2}^{D-1}\big(1-P_s^{\mathrm{pkt}}(i)\big). \label{eq:miss_deadline}
\end{equation}
For a stationary contention process ($P_s^{\mathrm{pkt}}(d)\equiv P_s^{\mathrm{pkt}}\,\forall d\ge 2$), \eqref{eq:miss_deadline} reduces to
\begin{align}
P_{>D}&=\big(1-P_s^{\mathrm{pkt}}\big)^{D-2}. \label{eq:stationary_miss}
\end{align}
In this work, we consider that the set of active LAPs and their transmission-policy distributions remain unchanged throughout an alarm event, while transmission patterns are independently sampled at each attempt. Accordingly, a retransmission may employ a different pattern without changing the underlying action-selection distribution. Under these assumptions, the per-slot success probability remains stationary across retransmissions.

Next, we define our optimization problem whose solution $\boldsymbol{\Psi}^*$ is a transmission policy that minimizes the deadline violation probability $P_{>D}$. Such optimization problem can be formulated as
\begin{subequations}\label{problem}
\begin{align}
\min_{\boldsymbol{\Psi}} \quad &P_{>D}, \label{eq:4a} \\
\text{subject to} \quad & \sum_{i=1}^{2^M} \psi_{n}(\grave{\boldsymbol{b_i}}) = 1, \quad \forall n\in \mathcal{\grave{N}}, \label{eq:4b} \\
& \boldsymbol{\Psi} \in [0,1]^{|\mathcal{N}| \times 2^M}. \label{eq:4c}
\end{align}
\end{subequations}

Since $\boldsymbol{\Psi}$  is fixed during an alarm event, $P_s^{\text{pkt}}(d)\equiv P_s^{\text{pkt}}(\Psi)$, and by~\eqref{eq:stationary_miss} $P_{>D}=(1-P_s^{\text{pkt}}(\Psi))^{D-2}$ is strictly decreasing in $P_s^{\text{pkt}}(\Psi)$ for $D\ge 3$. Hence, \eqref{problem} is equivalent to $\max_{\Psi} P_s^{\text{pkt}}(\Psi)$ subject to (7b) and (7c), and $D$ affects the value of $P_{>D}$ but not its minimizer. To solve the formulated problem, the complete information about the term ($\prod_{{n \in \mathcal{\grave{N}}}}p_{d_n}, \forall  \mathcal{\grave{N}}\in\mathcal{P}(\mathcal{N})$) is required, which is infeasible in practice as it requires knowing which LAPs are activated by the alarm event. Moreover, with $|\mathcal{\grave{N}}|>1$ and/or $|\mathcal{N}|>M$, the problem becomes NP-hard~\cite{NP}. Hence, we propose the following online-learning algorithm for solving the RA problem in~\eqref{problem}. 
    \begin{figure}[t] 
		\centering
		\includegraphics[width= 0.7\linewidth]{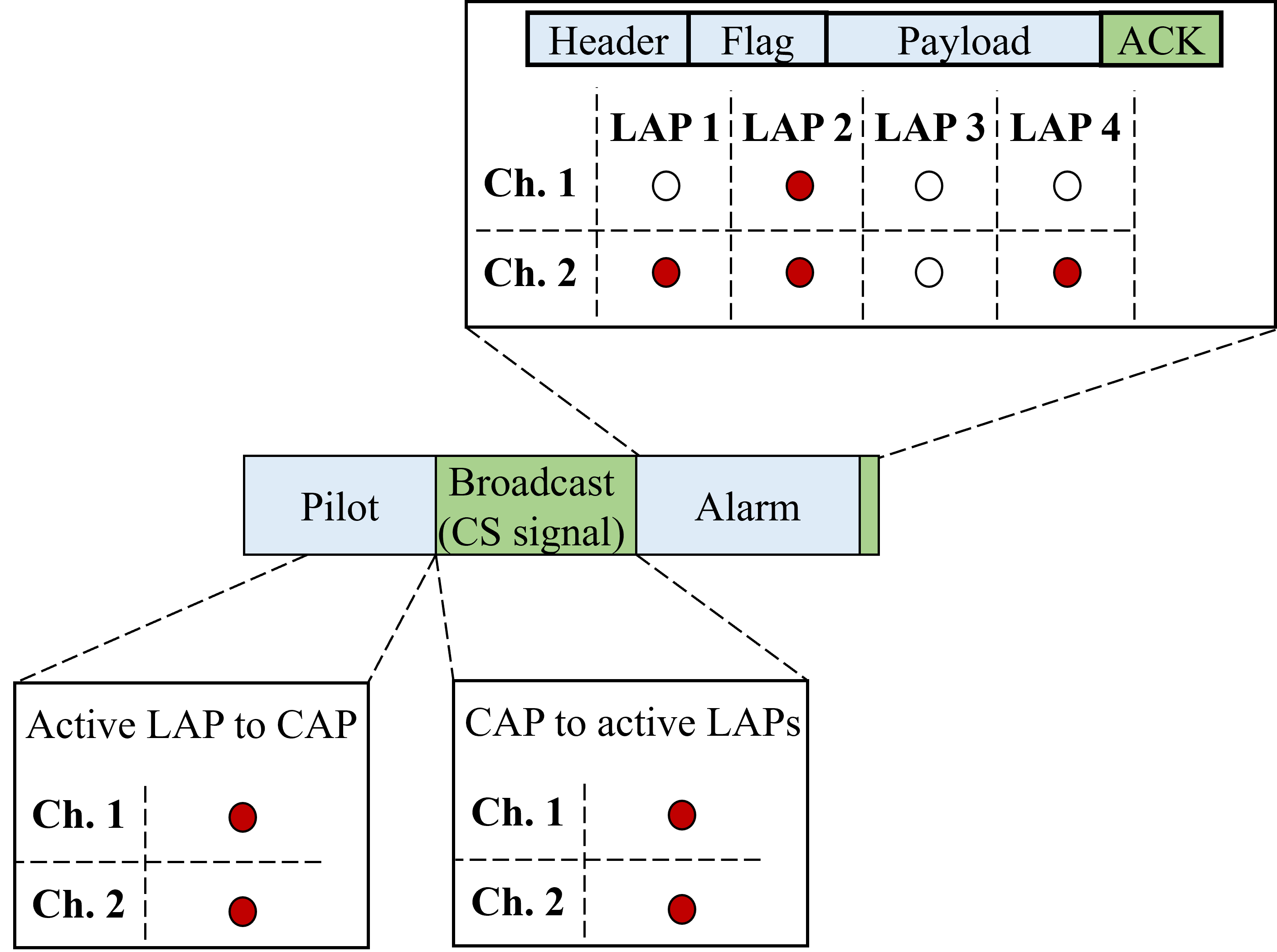}
		\caption{Illustration of the proposed protocol with $M=2$ and $|\mathcal{\grave{N}}|=4$. 
        \label{alarm}}
	\end{figure}
\section{The proposed deep contextual-Bandit RA Scheme}
\label{proposed}
In this work, we introduce a distributed multi-agent deep
contextual bandit, in which each active LAP $n$ observes a
context signal from the CAP, selects one of the
$2^M$ transmission patterns $\grave{\boldsymbol b}_n$ (the arms), and receives a
reward $r\in\{+1,-1\}$ depending on whether the alarm is delivered within the
deadline $D$. The transmission patterns are decided online and in a distributed fashion by the LAPs without explicit coordination overhead. Such decisions are updated at every alarm event based on the contention signature (CS), defined in the next subsection. The DNN utilizes the CS to select an action (transmission pattern) and receives a reward based on the transmission outcome. The proposed protocol is illustrated by Fig.~\ref{alarm}. The different components of the DNN are explained in the following subsections.
\subsection{The Contention Signature Signal}
While it is infeasible in practice to determine $|\mathcal{\grave{N}}|$, we introduce the CS signal, which implicitly informs LAP agents that a set of devices is contending (i.e., indicates the current contention level) which can be utilized (input to the DNN), in a distributed manner, to learn an optimal access configuration to minimize $P_{>D}$. The procedure for acquiring the CS signal is explained as follows. Each active LAP $n\in \mathcal{\grave{N}}$ first transmits a pilot signal to the CAP. Each pilot signal comprises $M$ symbols, each transmitted on a separate channel of the available $M$ channels. The CAP receives the superposition of the pilot signals from all active LAPs and broadcasts to the LAPs in the next time slot. The received signal by the LAPs is the CS signal which will be processed by the DNN during the online learning to determine the transmission pattern. The aggregated pilot signal $y$ received by the CAP can be expressed as
\begin{equation}\label{pilot}
\boldsymbol{y} = \sum_{n \in \mathcal{\grave{N}}} \sqrt{\rho} \operatorname{diag}(\boldsymbol{h}_n) \boldsymbol{x}_n + \bm{\sigma},
\end{equation}
where $\rho$ is the average transmit signal-to-noise ratio (SNR), $\boldsymbol{h}_n \in \mathbb{C}^{M\times 1}$ represents the $M$ channel coefficients between LAP $n$ and the CAP, $\boldsymbol{x}_n = [x_{1,n}, x_{2,n}, ..., x_{M,n}]^T\in \mathbb{C}^{M\times 1}$ is the pilot sequence from LAP $n$ and $\bm{\sigma}\sim\mathcal{CN}(\mathbf{0,I}_M)$ is the Additive White Gaussian Noise (AWGN) with normalized unit power. The CAP then transmits back the CS signal to all LAPs. The received CS signal $\boldsymbol{y}_n$ is given by
\begin{equation}\label{broadcast}
\boldsymbol{y}_n =  \sqrt{\rho} \operatorname{diag}(\boldsymbol{h}_n) \boldsymbol{y} + \bm{\hat{\sigma}},
\end{equation}
where $ \bm{\hat{\sigma}}\sim\mathcal{CN}(\mathbf{0,I}_M)$ is the normalized AWGN. For the sake of simplicity, we assume the same $\rho$ for the CAP transmission. The received CS provides information correlated with contention conditions.
\subsection{Reward and Action Selection}
 The action of an active agent (LAP) $n$ receiving the broadcast signal $\boldsymbol{y}_n$ is to select a transmission pattern $\boldsymbol{b}_n$. The overall system-level objective is to successfully transmit the alarm message  on at least one channel (i.e., $\delta(\mathcal{\grave{N}}, \boldsymbol{B})=1$) before the deadline $D$ expires. Upon a successful reception of an alarm message within the available $D-2$ slots, all active agents receive a shared ACK from the CAP and get a reward $r(\boldsymbol{b}_n)=1$. If no ACK is received within the available $D-2$ slots the agents get a reward $r(\boldsymbol{b}_n)=-1$.  The active agents attempt to maximize their expected reward by learning the
expected-reward function $V(\boldsymbol{y}_n,\grave{\boldsymbol{b}}_i)
=\mathbb{E}\big[r\mid\boldsymbol{y}_n,\grave{\boldsymbol{b}}_i\big]$,
$\forall i=1,\dots,2^M$, which represents the expected outcome of pattern
$\grave{\boldsymbol b}_i$ given the context $\boldsymbol{y}_n$. The LAP exploits the learned estimates through the $\epsilon$-greedy policy: it
selects
$\boldsymbol{b}_n=\arg\max_{\grave{\boldsymbol{b}}_i\in\{0,1\}^M}
V(\boldsymbol{y}_n,\grave{\boldsymbol{b}}_i)$ with probability $1-\epsilon$, or selects a
random pattern (exploration) with probability $\epsilon$. Due to fading, noise, and mobility, LAPs may observe different contention signatures even under similar contention conditions. Consequently, maintaining a separate action value for every possible context is impractical. We therefore use a parameterized
expected-reward function $\hat{V}(\boldsymbol{y}_n,\grave{\boldsymbol{b}}_i,\boldsymbol{w})$,
where $\boldsymbol{w}$ is the vector of tunable weights of the DNN layers, which is trained online by regressing $\hat{V}$ on the observed
rewards received from the CAP feedback. Finally, each LAP treats the patterns selected by the other
LAPs as part of its environment; this is a cooperative setting with a common
reward.
\subsection{Architecture and Training of the DNN}
 The input to the DNN is the received CS. This signal is then passed to two fully connected hidden layers, each with size $h$, followed by a dense output layer. A compact fully connected DNN is employed to approximate the expected reward associated with each transmission pattern. The number and size of hidden layers are selected empirically, and their impact on performance is evaluated in Section~\ref{results}. The DNN produces $2^M$ parameterized action values ($\hat{\boldsymbol{V}}$) at its output. To enhance training efficiency and stability, the DNN incorporates the ReLU activation function and the  Root Mean Square Propagation (RMSProp) optimizer. Each agent collects and store a tuple of $\{\boldsymbol{y_n}, \boldsymbol{b}_n,  r(\boldsymbol{b}_n)\}$ in its local memory with size $S$. Once the memory is full, the agent discards the oldest tuple to store a new one. In that way, the agents retain recent observations to train the DNN and perform fine-tuning of the weights without the need to store a full history in their limited memory. 
 
 The LAP agents follow an online training approach, by randomly sampling a mini-batch of size $B$ of the stored data and feeding it to the DNN to update the weight vector $\boldsymbol{w}$ by minimizing a cost function $J_n(\boldsymbol{w})$
 \begin{equation}
J_n(\boldsymbol{w}) = \frac{1}{B} \sum_{j=1}^{B} \left[ r_j(\boldsymbol{b}_n) - \boldsymbol{\hat{V_j}}(\boldsymbol{y_n}, \boldsymbol{b_n}, \boldsymbol{w}) \right]^2,
\end{equation}
where $r_j(\boldsymbol{b}_n)$ and $\hat{\boldsymbol{V_j}}$ denote the reward and the action-value function of the $j$-th sample, respectively. To prevent exploding gradients, we employ gradient norm clipping, which ensures that updates to the network weights remain within a stable range. The gradient vector $\nabla_w J(\boldsymbol{w})$ is clipped as
\begin{equation}
\Upsilon = \frac{\beta_0 \nabla_w J_n(\boldsymbol{w})}{\max(\|\nabla_w J_n(\boldsymbol{w})\|_2, \beta_0)},
\end{equation}
where $\beta_0$ is the threshold for $\|\nabla_w J_n(\boldsymbol{w})\|_2$. The random selection of mini-batches improves the model generalization and helps mitigate overfitting to specific short-term trends. Furthermore, training the DNN with mini-batches helps in reducing the variance in the DNN updates, leading to a more stable learning. It is worth noting that the selection of an appropriate mini-batch size $B$ is critical; a small $B$ can introduce high variance and cause oscillations in the learned protocol, while a large $B$ increases computational overhead and the risk of overfitting to the sampled mini-batch.
\begin{algorithm}[t]
\caption{Deep Contextual Bandit at Active Agent $n$}
\label{alg:nnbb}
\begin{algorithmic}[1]
\Require Context $\boldsymbol{y}_n$, $\epsilon$, $\boldsymbol{w}$, $D$
\State Evaluate $\hat{V}(\boldsymbol{y}_n,\grave{\boldsymbol{b}}_i,\boldsymbol{w})$ $\forall i=1,\dots,2^M$ using the DNN
\For{$d=2,\dots,D-1$}
  \State Draw $\varrho\sim\mathcal{U}(0,1)$
  \If{$\varrho>\epsilon$} $\boldsymbol{b}_n=\arg\max_{\grave{\boldsymbol{b}}_i\in\{0,1\}^M}\hat{V}(\boldsymbol{y}_n,\grave{\boldsymbol{b}}_i,\boldsymbol{w})$ 
  \Else{} select $\boldsymbol{b}_n$ uniformly at random
  \EndIf
  \State Transmit the alarm according to $\boldsymbol{b}_n$; observe the ACK
   \If{ACK is received} 
   \State $r(\boldsymbol{b}_n)=1$.
   \State \textbf{break}
   \Else{} $r(\boldsymbol{b}_n)=-1$.
   \EndIf
\EndFor
    \State Store $\{\boldsymbol{y}_n,\boldsymbol{b}_n,r(\boldsymbol{b}_n)\}$ in $S$
\State Sample a mini-batch of size $B$ from $S$
\State Update $\boldsymbol{w}$ with RMSProp
\State $\epsilon\gets\max(0.1,\epsilon-0.005)$
\end{algorithmic}
\end{algorithm}
\subsection{Slot-Level Operation of the Protocol}
The proposed RA protocol proceeds as follows.
\begin{itemize}
\item \textit{Slot 0 (pilot).} Each active LAP transmits its pilot sequence, one
symbol per channel. The CAP receives the aggregate signal $\boldsymbol{y}$ in~\eqref{pilot}.
\item \textit{Slot 1 (CS broadcast).} The CAP broadcasts the aggregate, and LAP
$n$ obtains $\boldsymbol{y}_n$ in~\eqref{broadcast}. The context is fixed for the rest of the
event, so the DNN is evaluated once.
\item \textit{Slots $d=2,\dots,D-1$ (alarm attempts).} In every attempt slot,
each active LAP draws a pattern $\boldsymbol{b}_n$ with the
$\epsilon$-greedy rule using $\hat{V}(\boldsymbol{y}_n,\grave{\boldsymbol{b}}_i,\boldsymbol{w})$ and
transmits the alarm on the channels with $b_{n,m}=1$. The CAP decodes the
alarm on a channel if and only if exactly one active LAP transmitted on it. The
alarm is delivered in slot $d$ if at least one such channel exists.
\item \textit{Feedback and retransmissions.}  The CAP broadcasts an
ACK to the active LAPs over the dedicated error-free control channel
if the alarm was decoded in slot $d$. The ACK is assumed
to arrive by the end of the same transmission slot and before the next slot starts. Once received, all
active LAPs stop and discard the alarm. Otherwise, if ACK is not received, each active LAP retransmits the alarm in slot $d+1$ by independently reselecting its transmission pattern using the same $\epsilon$-greedy policy. Specifically, the CS, DNN parameters, and exploration rate remain unchanged throughout an alarm event, whereas a new random draw is performed at each retransmission attempt. Consequently, the selected pattern may differ between consecutive attempts, potentially resolving previous collisions without additional signaling or DNN evaluation. Although reselection does not guarantee a different pattern, the nonzero exploration probability enables alternative transmission configurations to be selected. Under a fixed set of active LAPs and independent pattern sampling, the per-attempt success probability remains stationary, consistent with the deadline-violation analysis in Section~\ref{system-model}-C. The original CS does not need to be reacquired for every retransmission. This preserves the two-slot control overhead.
\item \textit{Online learning.} Following termination of the alarm event, each participating LAP updates its local DNN using mini-batches sampled from the collected experiences. The updated parameters are subsequently used for transmission-pattern selection during future alarm events. Algorithm~\ref{alg:nnbb} summarizes the operation of the deep contextual bandit learning.
\end{itemize}
\subsection{Computational Complexity} 
The computational complexity of the protocol is predominantly governed by three operations: (i) generation of action values using the DNN, (ii) selection of actions via an $\epsilon$-greedy strategy, and (iii) online updates during training based on the agent's local experience. Following the derivation in~\cite{complex}, the complexity of (i) and (iii) are $\zeta_1=\sum_{i=1}^{|\boldsymbol{l}|-1} l_{i+1} (2l_{i}+1)$ and $\zeta_3=B\zeta_1$, respectively. We have $\boldsymbol{l}=[l_1, l_2, ..., l_{|\boldsymbol{l}|}]^T$ with $l_1=|\boldsymbol{y}_n|$ (input layer) and $l_{|\boldsymbol{l}|}=|\grave{\boldsymbol{B}}|$ (output layer), while the remaining elements of $\boldsymbol{l}$ correspond to the size of the hidden layers. Considering a DNN with two hidden layers each with $h=1$ neuron, then we have $\boldsymbol{l}=[M, 1, 1, 2^M]$. For (ii), the action-selection complexity is bounded within the range $[3, 2+|\grave{\boldsymbol{B}}|]$. Accordingly, the lower bound $\zeta_{lb}$ of the protocol complexity can be given by~\cite{complex}
\begin{equation}\label{complexity_lb}
\begin{split}
\zeta_{lb} &= \zeta_1 + \zeta_3 + 3 \\
&= (B + 1)\zeta_1 + 3 \\
&= 90 \times 4^M + (123 + 60M) \times 2^M + 2M + 7.
\end{split}
\end{equation}
The upper bound complexity is given by
\begin{equation}\label{complexity_up}
\begin{split}
\zeta_{ub} &= \zeta_1 + \zeta_3 + 2 + |\grave{\boldsymbol{B}}|\\
&= \zeta_{lb} + 2^M - 1.
\end{split}
\end{equation}
From \eqref{complexity_lb} and \eqref{complexity_up}, the overall complexity is $\zeta={O}(4^M)$. The computational complexity of the proposed protocol is primarily governed by the number of orthogonal channels $M$. In dense in-X subnetworks, a small $M$ can severely impair system performance due to the increased likelihood of collisions among concurrently active devices. Conversely, increasing $M$ helps in reducing the deadline violation probability  by lowering contention, thereby enhancing overall performance. However, this improvement comes at the cost of a rapidly growing computational burden, specifically, an exponential increase in complexity with respect to $M$. As a result, deploying the protocol with a large $M$ may become impractical for resource-constrained LAPs, highlighting a fundamental trade-off between performance and deployment feasibility. One way to solve this is to divide the overall bandwidth into smaller subsets ($m_1, m_2, ..., m_M$) and assign a bandwidth subset to each group of LAPs. For instance, LAPs with high spatial correlation can be assigned different bandwidth subsets~\cite{spatial}.
\subsection{Adaptability to Network Dynamics}
The protocol requires neither knowledge of the number of active LAPs nor centralized coordination: each agent learns online from its local observations and the broadcast CS, which implicitly captures the current contention level. Hence, fluctuations in the set of active LAPs, channel conditions, or event-driven traffic bursts are reflected in the CS, and agents adapt their transmission decisions without reinitialization. This makes the scheme well suited to dynamic in-X subnetworks where network conditions, such as the set of active LAPs, channel quality, and interference levels, can change rapidly and unpredictably due to mobility, environmental variability, or sudden activation patterns triggered by alarm events.
\section{Performance Evaluation}
\label{results}
\begin{table}[t]
\centering
\caption{Summary of system-level simulation parameters}
\label{tab:sys_sim_params}
\renewcommand{\arraystretch}{1.08}
\setlength{\tabcolsep}{6pt}
\begin{tabular}{|l|l|}
\hline
\textbf{Parameter} & \textbf{Value} \\
\hline
Deployment area & $50$m $\times$ $50$m\\
Transmit power per channel & -10 dBm\\
Velocity of a subnetwork ($v$) & $2$ m/s\\

Time slot duration ($T_C$) & \num {10} ms\\
Deadline threshold $(D)$ & \num {20} time slots~\cite{deadline_value}\\
spatial attenuation rate ($\eta$) &  0.6\\
Number of hidden layers & 2\\
Neurons per hidden layer $h$ & 1\\
Mini-batch size $B$ & $2^M \times 30$ \\
Buffer size $S$ & $2^M \times 100$~\cite{size} \\
\hline
\end{tabular}
\end{table}
In this section, we evaluate the performance of the proposed RA method via comprehensive MATLAB simulations using the parameters listed in Table~\ref{tab:sys_sim_params} (unless stated otherwise).
\subsection{Simulation Setup}
Our evaluation focuses on a single Z-subnet comprising $N$ L-subnets (LAPs). The L-subnets are randomly distributed in a 50 m $\times$ 50 m deployment area, which is a typical scenario in industrial subnetworks~\cite{InF, par1}. A snapshot-based mobility model is adopted. At each snapshot, the subnetworks are randomly placed within a rectangular deployment region following a uniform spatial distribution. Each subnetwork then proceeds to move in a randomly selected direction at a constant speed $v$. The direction of motion is updated whenever a subnetwork either reaches the boundary of the area or comes within $1.5$ meters of another subnetwork, where this minimum separation constraint is enforced to prevent unrealistic overlaps or collisions. Mobility of the L-subnets introduces time-varying contention conditions, as the movement changes their distances from the event epicenter and, consequently, their probabilities of detecting the same alarm event. Mobility may also affect the wireless channels between the LAPs and the CAP, resulting in variations in the received CS signals, requiring the LAPs to adapt their transmission strategies to changing network conditions. The learning phase of the DNN starts with a greater emphasis on exploration than exploitation with $\epsilon$ gradually decreasing from 1 to 0.1 with a step of 0.005. The learning rate in RMSProp is a hyperparameter and is gradually reduced by a decay rate of 0.015 following each alarm event.

Our proposed RA protocol is compared with two benchmarks. The first one is random channel selection (RCH) where each LAP randomly selects a transmission pattern. The second one is a multi-agent RA protocol, referred to as MAP-RA. We implement MAP-RA as a myopic Q-learning with linear approximation of the action-value function, enabling the algorithm to accommodate a continuous state space without explicitly storing action values for every state-action pair. For an active LAP $n$, the estimated action value  $Q_1(\boldsymbol{y}_n,\grave{\boldsymbol{b}}_i,
\boldsymbol{\theta}_n)$ is expressed as
\begin{equation}
Q_1(\boldsymbol{y}_n,\grave{\boldsymbol{b}}_i,
\boldsymbol{\theta}_n)
=
\boldsymbol{\theta}_n^{T}
\boldsymbol{\phi}(\boldsymbol{y}_n,
\grave{\boldsymbol{b}}_i),
\label{eq:mqlfa_q}
\end{equation}
where $\boldsymbol{\theta}_n\in
\mathbb{R}^{2M\times 1}$ denotes the weight vector maintained locally by each LAP $n$ and $\boldsymbol{\phi}(\boldsymbol{y}_n,
\grave{\boldsymbol{b}}_i)=[\operatorname{diag}(\tilde{\boldsymbol{y}}_n
\tilde{\boldsymbol{y}}_n^{H}),\grave{\boldsymbol{b}}_i]^T \in\mathbb{R}^{2M\times 1}$. $\tilde{\boldsymbol{y}}_n$ is the normalized $\boldsymbol{y}_n$ and is given as
\begin{equation}
\tilde{\boldsymbol{y}}_n =
\frac{
\overbrace{
\boldsymbol{y}_n
- \min\left(\operatorname{Re}(\boldsymbol{y}_n)\right)
- \min\left(\operatorname{Im}(\boldsymbol{y}_n)\right)j
}^{\boldsymbol{\kappa}}
}{
\max\left(\operatorname{abs}(\boldsymbol{\kappa})\right)
}.
\end{equation}
In MAP-RA, upon detecting an alarm, each activated LAP first transmits its pilot sequence and subsequently obtains the context $\boldsymbol{y}_n$. Using this observation, the LAP evaluates the estimated action values for all $2^M$ candidate transmission patterns according to \eqref{eq:mqlfa_q}. The transmission pattern is then determined through an $\epsilon$-greedy selection policy, following the same exploration-rate schedule adopted by our proposed RA protocol.
After executing the selected transmission pattern $\boldsymbol{b}_n$, the device obtains the corresponding reward $r(\boldsymbol{b}_n)$ from the receiver feedback. Its local weight vector is subsequently adjusted according to
\begin{equation}
\boldsymbol{\theta}_n
\leftarrow\;
\boldsymbol{\theta}_n
+\tau
\Big(
r(\boldsymbol{b}_n)
-Q_1(\boldsymbol{y}_n,\boldsymbol{b}_n,
\boldsymbol{\theta}_n)
\Big)
\boldsymbol{\phi}(\boldsymbol{y}_n,
\boldsymbol{b}_n),
\label{eq:mqlfa_update}
\end{equation}
where $\tau$ is the learning rate. Since MAP-RA incorporates the CS into its action-value estimation, it provides a suitable baseline for the proposed scheme.

We evaluate the performance of each scheme based on the probability $P_{\le D}$, referred to in the results as the probability of in-time alarm delivery. We consider an alarm scenario as in~\cite{deadline_value} with a deadline of $200$ ms. The reported results are averaged over $100$ simulation runs, each with a length of $10^5$ time slots. 
\subsection{Results and Discussion}
    \begin{figure}[t] 
		\centering
		\includegraphics[width= 1\linewidth]{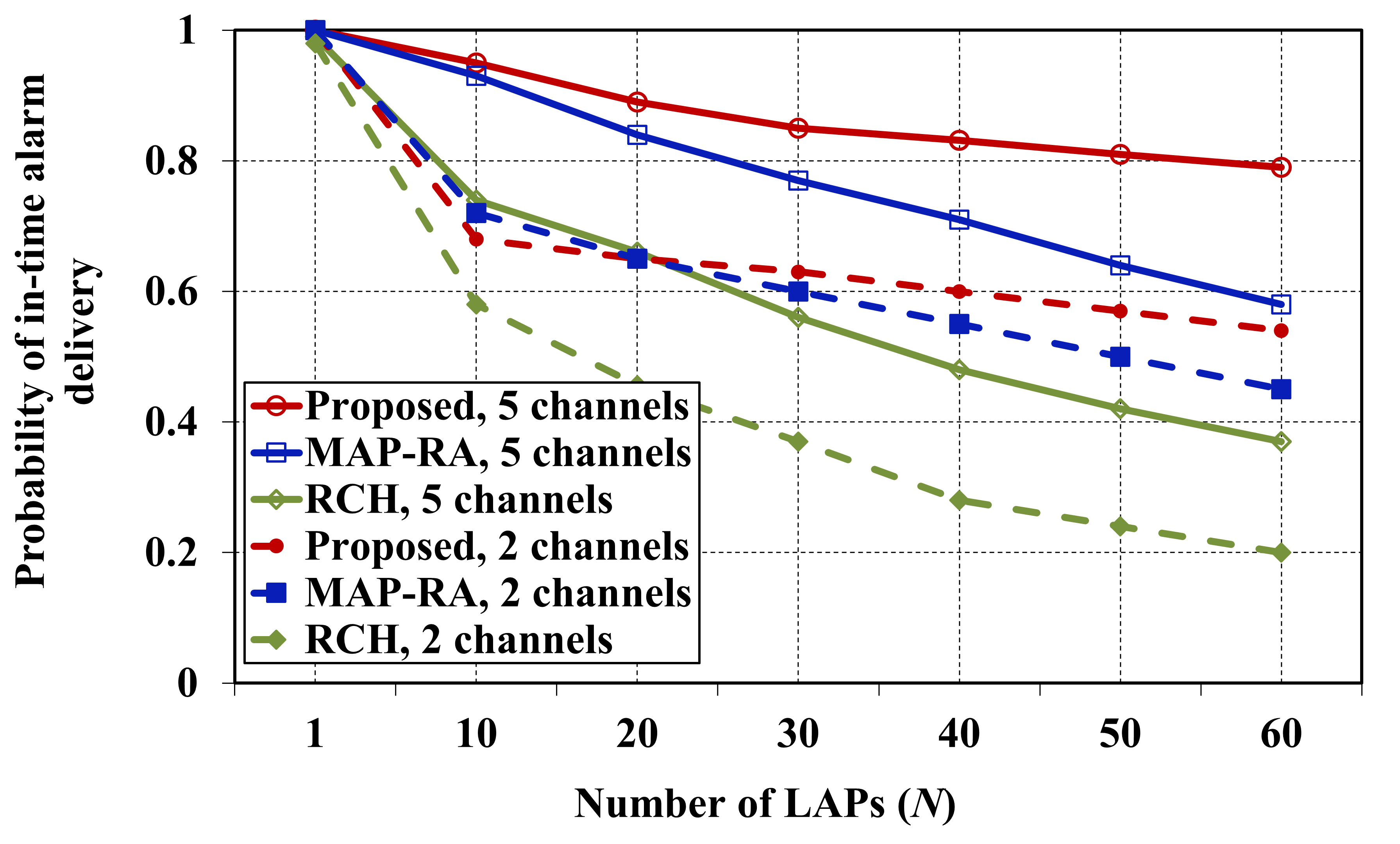}
		\caption{Performance comparison of probability of in-time alarm delivery for different numbers of L-subnets ($N$) with $M=\{2, 5\}$.  \label{LAPs}}
	\end{figure}
Figure~\ref{LAPs} shows the probability of in-time alarm delivery $P_{\le D}=1-P_{>D}$, as a function of the number of LAPs. As the network size increases, the performance of all evaluated protocols gradually deteriorates. This behavior is expected, since a larger number of LAPs leads to stronger contention for the shared channel resources, thereby increasing the probability of collisions and delayed transmissions. Despite this general degradation trend, our proposed scheme consistently achieves the highest $P_{\le D}$ across all network sizes. The MAP-RA protocol utilizes the contention signature, however, our proposed mechanism introduces the DNN which  processes the CS more effectively leading to better action values than MAP-RA. On the other hand, the RCH protocol shows the most significant degradation due to its limited adaptability to increasing contention levels. In contrast, the proposed approach enables distributed LAPs to learn context-aware transmission policies that effectively mitigate contention and improve the likelihood of delivering alarms within strict delay constraints. As a result, the performance advantage of the proposed scheme becomes increasingly evident as the network size grows. For instance, when $N=30$ and $M=5$, the proposed method achieves approximately $8\%$ ($30\%$) higher $P_{\le D}$ compared to MAP-RA (RCH), while the performance gain increases to $21\%\,(42\%)$ when $N=60$. These results highlight the superior scalability and robustness of the proposed approach in dense industrial scenarios. Particularly, our proposed method can support dense deployment of L-subnets to enhance the coverage (event-detectability) and minimize blind spots, while maintaining timely delivery of emergency alarms.  
    \begin{figure}[t] 
		\centering
		\includegraphics[width= 1\linewidth]{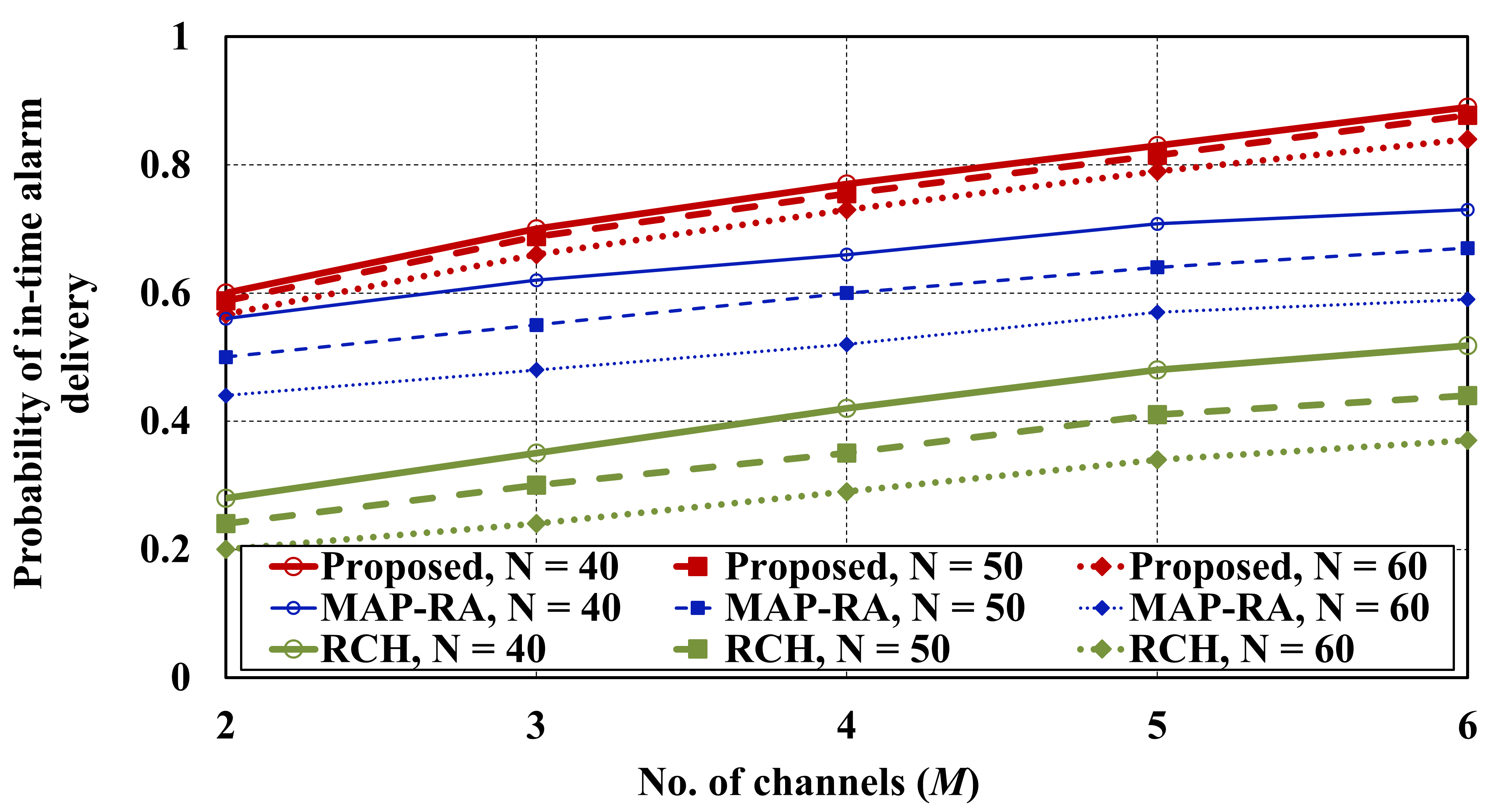}
		\caption{Performance comparison of probability of in-time alarm delivery for different numbers of channels $(M)$ with $N=\{40, 50, 60\}$.  \label{channels}}
	\end{figure}

In Fig.~\ref{channels}, we show the impact of the number of available channels $M$ on $P_{\le D}$ under different network densities. As expected, increasing $M$ improves the performance of all considered schemes as additional channels alleviate channel contention among LAPs, thereby increasing the likelihood that alarm packets are delivered within the required delay constraint.  Nevertheless, the proposed approach achieves a significantly higher $P_{\le D}$ for all values of $M$, particularly in scenarios with limited number of channels (e.g., $M=2$). In such cases, efficient coordination among competing LAPs becomes crucial, and the proposed method effectively learns transmission behaviors that minimize collisions. This capability stems from the deep learning model’s ability to capture and exploit patterns in the contention dynamics, allowing LAPs to adapt their access decisions even under severe channel scarcity. A quantitative example can be observed for the dense network case with $N=60$, where the proposed method gains a $28\%$ improvement in $P_{\le D}$ when $M$ increases from 2 to 5, whereas MAP-RA achieves only a $13\%$ improvement over the same range. Another notable observation from this figure is the robustness of the proposed scheme to network densification. While the performance of MAP-RA and RCH noticeably deteriorates as the number of LAPs increases (from $N=40$ to $N=60$), the proposed method maintains maintains a relatively stable performance across different network sizes, highlighting its effectiveness in dense alarm-reporting scenarios.
    \begin{figure}[t] 
		\centering
		\includegraphics[width= 1\linewidth]{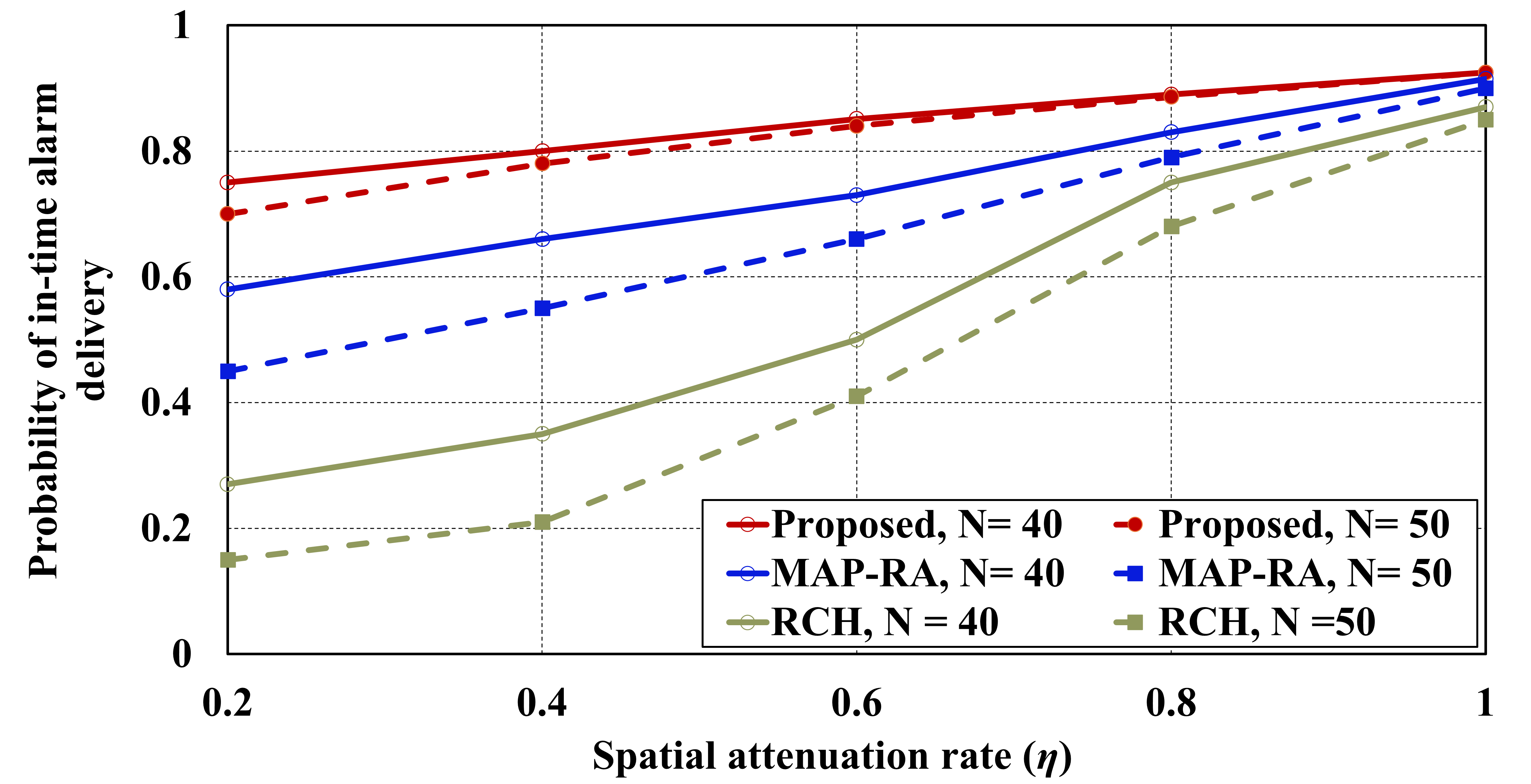}
		\caption{Performance comparison of probability of in-time alarm delivery for different values of the spatial attenuation rate $(\eta)$ with $N=\{40, 50\}$ and $M=5$.  \label{activation}}
	\end{figure}

Figure~\ref{activation} illustrates the effect of the spatial attenuation rate $\eta$ on the performance of the three methods. As $\eta$ decreases, the activation probability of the LAPs increases, leading to increased collision probability  which degrades the  $P_{\le D}$ for all methods. However, the proposed method experiences a lower rate of performance degradation compared to MAP-RA and RCH. For low values of $\eta$, e.g., 0.2, the values of $P_{\le D}$ of MAP-RA and RCH fall below 0.5 with $N=50$, while that of the proposed scheme remains significantly higher~($>$0.7). Therefore, the proposed method can enhance the network coverage, while maintaining reliable timely delivery of alarm events. 

    \begin{figure}[t] 
		\centering
		\includegraphics[width= 0.9\linewidth]{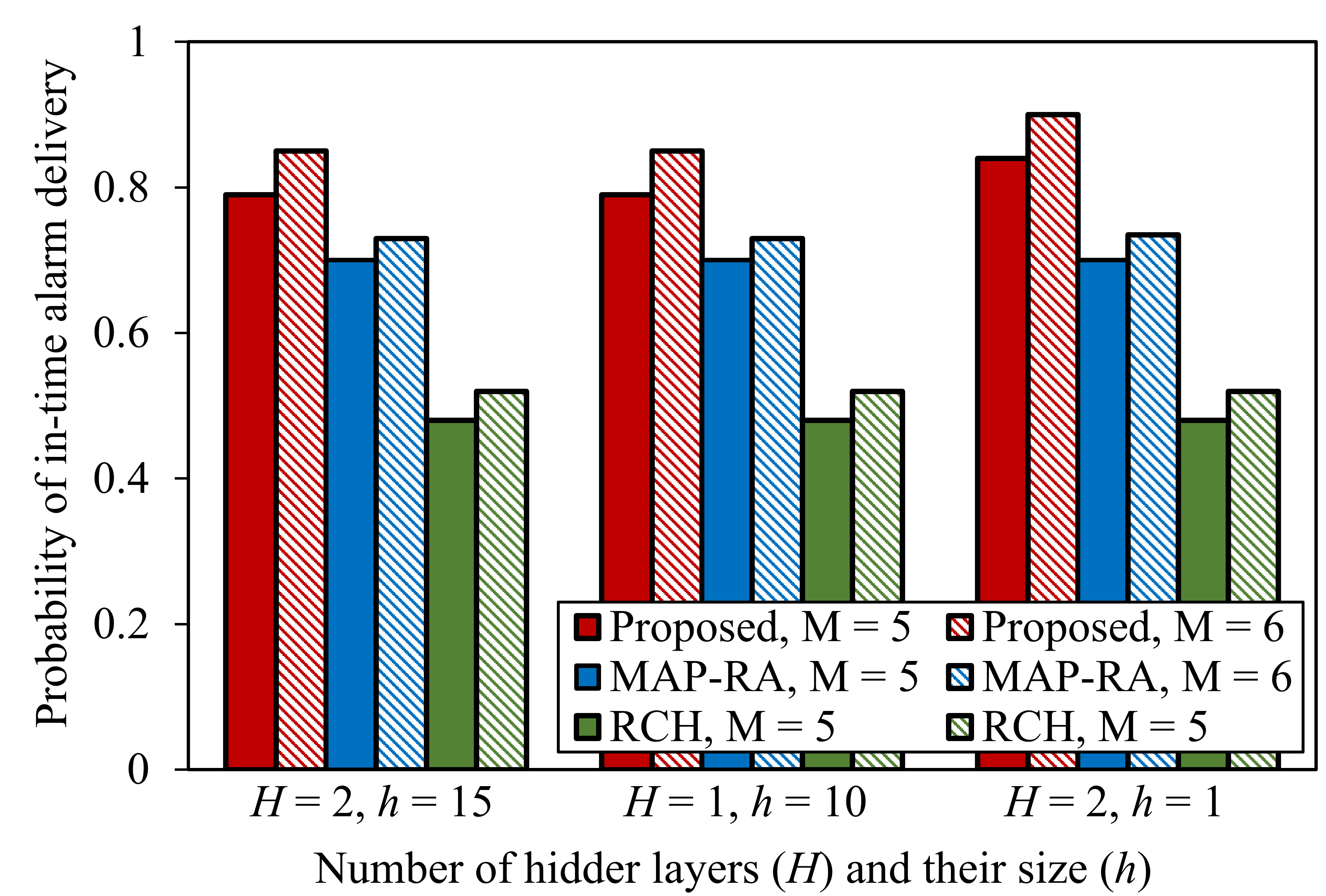}
		\caption{Performance comparison of probability of in-time alarm delivery for different DNN configurations with $N=40$ and $M=\{5, 6\}$.  \label{layers1}}
	\end{figure}
\begin{figure}[t!] 
		\centering
		\includegraphics[width= 1\linewidth]{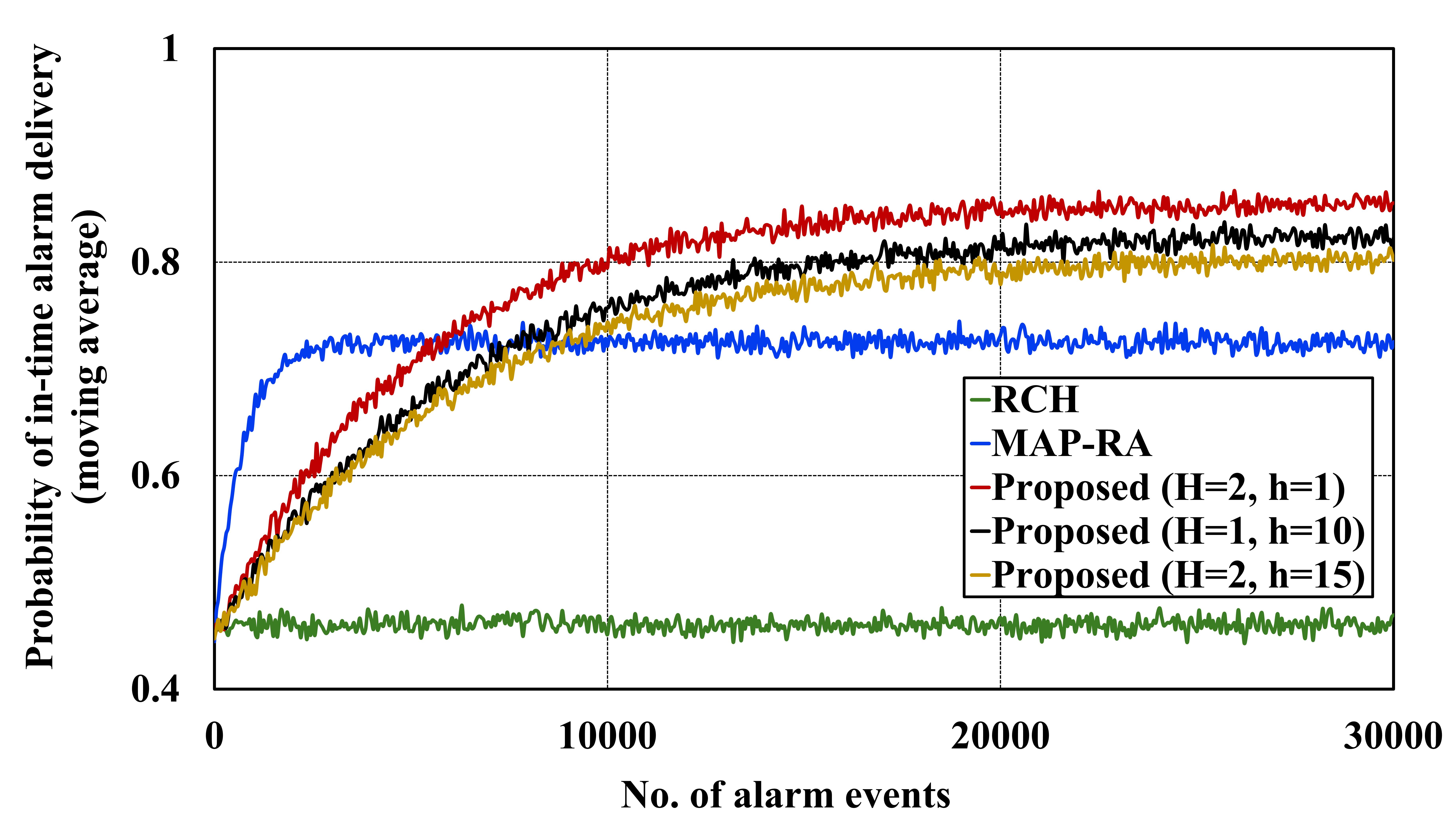}
		\caption{Performance comparison during training for different DNN architectures with $N=40$ and $M=5$.  \label{layers}}
	\end{figure}

In Fig.~\ref{layers1}, we evaluate the performance as a function of the neural network architecture, specifically the number of hidden layers $H$ and the number of neurons per layer $h$. The figure shows that the proposed method consistently achieves higher $P_{\le D}$ than both baseline approaches across all tested neural network configurations. Another interesting observation  is that the configuration with the smallest hidden layer size ($H=2, h=1$) yields the best overall performance among the considered architectures. Moreover, according to the results reported in Fig.~\ref{layers}, this configuration outperforms the  baseline approaches throughout the training phase. The superior performance of the compact architecture suggests that increasing the model capacity does not necessarily improve learning in the considered setting. Larger DNNs may require additional training samples or more careful hyperparameter tuning to achieve comparable performance. Moreover, the compact architecture reduces computational complexity, making it attractive for resource-constrained LAPs.
\section{Conclusion}\label{sec:conclusions}
In this work, we have proposed a deep contextual-bandit-based RA protocol for timely alarm delivery in 6G industrial in-X subnetworks. Each LAP processes a broadcast contention signature with a lightweight DNN and selects its transmission pattern via an $\epsilon$-greedy policy, refining it online from ACK-based rewards; this achieves implicit coordination among LAPs without centralized control or knowledge of the number of active devices. We further showed, via a DTMC analysis, that minimizing the deadline-violation probability is equivalent to maximizing the per-slot success probability. Simulation results show that the proposed scheme achieves a higher in-time delivery probability than the benchmark schemes, with a gain over MAP-RA that grows with network density, reaching 21\% with 60 subnetworks, and it is the most robust to channel scarcity and spatially widespread events. Moreover, a compact architecture with two single-neuron hidden layers achieved the best performance, making the scheme suitable for resource-constrained LAPs. Future work will extend the framework to heterogeneous (periodic and event-driven) traffic, include channel impairments such as fading and shadowing in the analysis, and explore federated learning to enhance coordination while preserving distributed operation.

	\bibliographystyle{IEEEtran}
\bibliography{main}
	
\end{document}